\documentclass[journal]{IEEEtran}

\ifCLASSINFOpdf
\else
   \usepackage[dvips]{graphicx}
\fi
\usepackage{url}

\makeatletter

\newcommand{\Rmnum}[1]{\expandafter\@slowromancap\romannumeral #1@}
\makeatother

\usepackage{graphicx}

\usepackage[cmex10]{amsmath}
\usepackage{array}

\usepackage{cases} 
\usepackage{bm} 

\begin{document}

\title{Inverse LDM$^T$ and LU Factorizations of a Partitioned Matrix with
the Square-root and Division Free Version for V-BLAST}

\author{Hufei~Zhu
\thanks{H. Zhu is with the College of Computer Science and Software, Shenzhen University, Shenzhen 518060, China (e-mail:
zhuhufei@szu.edu.cn).  This paper was presented
in part at the IEEE Vehicular Technology Conference (VTC 2010 Fall), 6-9 Sept., 2010.}
}

\markboth{Journal of \LaTeX\ Class Files, Vol. 14, No. 8, August 2015}
{Shell \MakeLowercase{\textit{et al.}}: Bare Demo of IEEEtran.cls for IEEE Journals}
\maketitle

\begin{abstract}
This letter proposes the inverse LDM$^T$ and LU factorizations
of a matrix partitioned into $2 \times 2$ blocks, which include the square-root and division free version.
The proposed square-root and division free inverse LDM$^T$ factorization
is  applied to compute the initial
   estimation error covariance matrix
    $\mathbf{Q}$ for the recursive  V-BLAST algorithm, which can save $K-1$ divisions (where $K$ is the number of transmit antennas), and
    requires about the  same computational complexity
    as the corresponding algorithm to compute $\mathbf{Q}$ in the existing recursive V-BLAST
    algorithm~\cite{TransWC2009ReCursiveBlast,zhfICC2009}.
 The proposed square-root and division free inverse LDM$^T$ factorization can also be
     applied to propose the square-root and division free implementation for the square-root V-BLAST algorithm in \cite{zhfvtc2008}, where the
 wide-sense Givens rotation in \cite{zhfVTC2010DivFree} is utilized.
  With respect to the  existing square-root V-BLAST algorithms~\cite{zhfvtc2008,BLASTtransSP2015}, the proposed
  square-root and division free V-BLAST algorithm requires about the same computational complexity, and can avoid the
  square-root and division operations.
\end{abstract}


\IEEEpeerreviewmaketitle

\section{Introduction}

%

\IEEEPARstart{M}{ultiple}-input multiple output (MIMO) wireless communication systems can achieve
 very high spectral efficiency in rich multi-path environments~\cite{MIMOintroduce}.
Bell Labs Layered Space-Time architecture (BLAST), including the relative simple vertical
BLAST (V-BLAST),  is such a  system that transmits independent data streams simultaneously from multiple antennas~\cite{zhf1}
to maximize the data rate.
 V-BLAST usually utilizes the ordered successive interference cancellation (OSIC) detector~\cite{zhf1},
 to detect the data streams iteratively with the optimal ordering.
 In each iteration, the OSIC detector utilizes a zero-forcing (ZF) or minimum mean-square error (MMSE) filter
 to detect the data stream with the highest
signal-to-noise ratio (SNR) among all undetected data streams,
and then subtracts
the effect of the detected
data stream in the received signal vector.

%

The OSIC detector for V-BLAST requires high computational complexity.
Thus fast algorithms have been proposed for V-BLAST
\cite{zhf2}-\cite{zhfICC2009},
of which typical examples are
the  square-root algorithms
\cite{zhf2}-\cite{BLASTtransSP2015}
 and
 the recursive algorithms \cite{zhf3}-\cite{zhfICC2009}.
In the
OSIC detection phase,
the recursive V-BLAST algorithms update  the
   estimation error covariance matrix
    $\mathbf{Q}={\mathbf{R}}^{-1}$  recursively, while
the square-root V-BLAST algorithms update the square-root of $\mathbf{Q}$, i.e.,
${\bf{\Gamma }}$ satisfying  ${\bf{\Gamma }} {\bf{\Gamma }}^H = {\bf{Q}}$.

In fixed-point implementations, it is highly desirable to avoid square-root
and division operations, since they
are demanding in terms of the required bit precision and clock
cycles \cite{WCNCcholesky,kjLiuGIVENS}.
To compute the initial
square-root ${\bf{\Gamma }}$, the inverse  Cholesky factorization in \cite{zhfvtc2008} and the corresponding algorithm in
\cite{BLASTtransSP2015} (by the Cholesky factorization and the back-substitution~\cite{Matrix_Computations_book}) both reduce half divisions.
 Moreover,
to implement a spherical MIMO detector,
the alternative Cholesky factorization was proposed in
\cite{WCNCcholesky}
 to avoid both square-root and division
operations.    This letter makes progress along this direction, and gives an efficient square-root and division free algorithm
to compute the initial square-root
 for the square-root V-BLAST algorithm, which can  also be utilized to compute the initial $\mathbf{Q}$ for the recursive V-BLAST algorithm.

In Matlab,
  the  ``inv" function~\cite{Matlab_inv_function_introduce} for the matrix inversion computes the $\mathbf{LU}$ factors of a matrix $\mathbf{R}$,  inverts the $\mathbf{LU}$ factors,  and multiplies the inverses
  to obtain $\mathbf{Q}=\mathbf{R}^{-1}$.
  When
   $\mathbf{R}$ is Hermitian, the $\mathbf{LU}$ factors become the ${\mathbf{LDL}}^T$ factors~\cite{Matlab_inv_function_introduce}.
This letter
proposes the improved implementations for the inversion algorithm in the Matlab  ``inv" function, which can be utilized in V-BLAST.
We propose an efficient inverse ${\mathbf{LDM}}^T$ factorization to compute the $(k+i)^{th}$ order inverse ${\mathbf{LDM}}^T$ factors
from the $k^{th}$ order inverse ${\mathbf{LDM}}^T$ factors by just one iteration, where $k$ and  $i$ can be any positive integers.
Then the proposed inverse ${\mathbf{LDM}}^T$ factorization is transformed into the proposed inverse ${\mathbf{LU}}^T$ factorization.
Moreover, from the proposed inverse ${\mathbf{LDM}}^T$ factorization, we deduce the alternative division-free
inverse ${\mathbf{LDM}}^T$ factorization,
which is utilized by the recursive V-BLAST algorithm~\cite{TransWC2009ReCursiveBlast,zhfICC2009}
to compute the initial ${\mathbf{Q}}$, and is also applied to develop a full square-root and division free implementation
of the square-root V-BLAST algorithm in \cite{zhfvtc2008}.

The V-BLAST System model is overviewed in Section \Rmnum{2}. 
Section \Rmnum{3} proposes the inverse LDM$^T$ and LU factorizations with
 the square-root and division free version.
Then the proposed square-root and division free inverse LDM$^T$ factorization
is applied in V-BLAST in Section \Rmnum{4}.
The complexity of the presented algorithms is
evaluated in Section \Rmnum{5}. Finally, we make conclusion in
Section \Rmnum{6}.


%
%

\section{V-BLAST System Model}


The considered V-BLAST system consists of $K$ transmit antennas and
$N(\ge K)$ receive antennas in a rich-scattering and flat-fading
wireless channel. At the transmitter, the data stream is
de-multiplexed into $K$ sub-streams. Then each sub-stream is encoded
and fed to its respective transmit antenna. 
Let ${\bf{s}}=
[s_1 ,s_2 , \cdots ,s_K ]^T$ denote the vector of transmit symbols
from $K$ antennas, and assume $ E({\bf{ss}}^H ) = \sigma _{{s}}^2
{\bf{{\rm I}}}_K $ where $ {\bf{{I}}}_K $ is the identity matrix with size $K$.
Then the received symbol vector is 
\begin{equation}\label{equ:1}
{\bf{x}} = {\bf{H}} \cdot {\bf{s}} + {\bf{n}},
\end{equation}
where ${\bf{n}}$ is the $N\times 1$ complex Gaussian noise vector
with zero mean and covariance $\sigma _n ^2 {\bf{{\rm I}}}_N$, and
${\bf{H}}$ is the
$N\times K$ complex channel matrix with statistically independent
entries.

The minimum mean-square error (MMSE) detection of $\bf{s}$ is
 \begin{equation}\label{equ:2}
{\bf{\hat s}} = \left( {{\bf{H}}^H {\bf{H}} + \alpha {\bf{I}}_K }
\right)^{ - 1} {\bf{H}}^H {\bf{x}},
\end{equation}
where $\alpha  = \sigma _n^2 /\sigma _s^2$,
$(\bullet
)^{-1}$ and $ ( \bullet )^H $ denote matrix
inversion and matrix conjugate transposition,
respectively. Let
 \begin{equation}\label{S28R}
{\bf{R}} = {\bf{H}}^H  \cdot {\bf{H}} + \alpha {\bf{I}}_K.
\end{equation}
The estimation error covariance matrix  is \cite{zhf2}
 \begin{equation}\label{equ:4}
 {\bf{Q}} = {\bf{R}}^{ - 1}=\left( {{\bf{H}}^H {\bf{H}} + \alpha {\bf{I}}_K }
\right)^{ - 1},
\end{equation}
and the square-root of ${\bf{Q}}$  is
${\bf{\Gamma }}$ satisfying
\begin{equation}\label{squareRootDefine32198}
 {\bf{\Gamma }} {\bf{\Gamma }}^H = {\bf{Q}}.
\end{equation}


The conventional V-BLAST detects $K$ entries of ${\bf{s}}$ by $K$
iterations with the optimal ordering. In the $k^{th}$ ($k=K,K-1,\cdots,2$) iteration, the entry
with the highest post detection signal-to-noise ratio (SNR) among
all the undetected $k$ entries is detected by a linear MMSE or
zero-forcing (ZF) filter. Then its effect is subtracted from the
received symbol vector \cite{zhf1,zhf2}, and accordingly ${\bf{Q}}_{|(k-1)}$
or ${\bf{\Gamma}}_{|(k-1)}$ corresponding to all the undetected $k-1$ entries
needs to be computed.
%

The first $k$ columns of ${\bf{H}}$ can be represented as
\begin{equation}\label{equ:10}
{\bf{H}}_k  = [{\bf{h}}_1 ,{\bf{h}}_2, ...,{\bf{h}}_k ],
\end{equation}
where  ${\bf{h}}_m$ ($1\le m \le k$) denotes the $m^{th}$ column of
${\bf{H}}$.
Then ${\bf{R}}_k$ and ${\bf{Q}}_k$ are defined from ${\bf{H}}_{k}$
by (\ref{S28R}) and (\ref{equ:4}), respectively, while ${\bf{R}}_{k+1}$
satisfies \cite{zhf3}
 \begin{multline}\label{equ:11}
{\bf{R}}_{k+1} = \\
\left[ {\begin{array}{*{20}c}
   {{\bf{R}}_{k} } & {\bf{H}}_{k}^H {\bf{h}}_{k+1}  \\
   {\bf{h}}_{k+1}^H {\bf{H}}_{k}  & {\bf{h}}_{k+1}^H  {\bf{h}}_{k+1} + \alpha  \\
\end{array}} \right] = \left[ {\begin{array}{*{20}c}
   {{\bf{R}}_{k} } & {{\bf{v}}}  \\
   {{\bf{v}}^H } & t  \\
\end{array}} \right].
\end{multline}

\section{Efficient Inverse LDM$^T$ and LU Factorizations of a Partitioned Matrix}


Assume that $i$ rows and columns are added to a $k \times k$ general square matrix
${{\mathbf{R}}_{k}}$ to form a $(k+i) \times (k+i)$  matrix ${{\mathbf{R}}_{k+i}}$,
 which is written as a matrix partitioned into $2 \times 2$ blocks,   i.e.,
    \begin{equation}\label{R2RVT324913}
\mathbf{R}_{k+i}^{{}}=\left[ \begin{matrix}
   {{\mathbf{R}}_{k}} & \mathbf{V}  \\
   {{\mathbf{Y}}^{H}} & \mathbf{T}  \\
\end{matrix} \right].
 \end{equation}
Obviously (\ref{equ:11}) is a special case of
(\ref{R2RVT324913}) with $i=1$ and the Hermitian $\mathbf{R}_{k+i}$.
 The upper-triangular LDM$^T$ factors of ${{\mathbf{R}}}^{-1}={{\mathbf{Q}}}$ satisfy
   \begin{equation}\label{LDMtFactorsDefine398da32}
\mathbf{LD}{{\mathbf{M}}^{H}}={{\mathbf{R}}^{-1}}={{\mathbf{Q}}},
 \end{equation}
from which
  we can deduce
   \begin{equation}\label{MinvDL2Re4332}
{{\mathbf{M}}^{-H}}{{\mathbf{D}}^{-1}}{{\mathbf{L}}^{-1}}=\mathbf{R},
 \end{equation}
 where
   ${{\mathbf{M}}^{-H}}$
 and ${{\mathbf{L}}^{-H}}$ are the conventional
 lower-triangular  LDM$^T$ factors~\cite{Matrix_Computations_book} of ${{\mathbf{R}}}$.

\subsection{Inverse LDM$^T$ Factorization of a Partitioned Matrix}

Obviously the upper-triangular LDM$^T$ factors of ${{\mathbf{R}}_{k+i}^{-1}}$ satisfy
 \begin{subnumcases}{\label{LDMdefineAll313943}}
{{\mathbf{L}}_{k+i}}\text{=}\left[ \begin{matrix}
   {{\mathbf{L}}_{k}} & \mathbf{A}  \\
   \mathbf{0} & \mathbf{F}  \\
\end{matrix} \right]  &  \label{L2LAF32413}\\
{{\mathbf{D}}_{k+i}}=\left[ \begin{matrix}
   {{\mathbf{D}}_{k}} & \mathbf{0}  \\
   \mathbf{0} & \mathbf{G}  \\
\end{matrix} \right]   &  \label{D2DG3243} \\
{{\mathbf{M}}_{k+i}}\text{=}\left[ \begin{matrix}
   {{\mathbf{M}}_{k}} & \mathbf{B}  \\
   \mathbf{0} & \mathbf{E}  \\
\end{matrix} \right].    &  \label{M2MBE2314}
\end{subnumcases}
In (\ref{LDMdefineAll313943}), $\mathbf{A}$, $\mathbf{F}$, $\mathbf{G}$, $\mathbf{B}$ and $\mathbf{E}$
can be computed by
\begin{subnumcases}{\label{ABFGEcomputeE84r9834}}
\mathbf{A}=-\mathbf{L}_{k}^{{}}\mathbf{D}_{k}^{{}}\mathbf{M}_{k}^{H}\mathbf{VF} &  \label{A2LFMVF8430}\\
 {{\mathbf{B}}^{H}}=-{{\mathbf{E}}^{H}}{{\mathbf{Y}}^{H}}\mathbf{L}_{k}^{{}}\mathbf{D}_{k}^{{}}\mathbf{M}_{k}^{H} &  \label{B2EYLDM3243}\\
\mathbf{FG}{{\mathbf{E}}^{H}}={{(\mathbf{T}-{{\mathbf{Y}}^{H}}\mathbf{L}_{k}^{{}}\mathbf{D}_{k}^{{}}\mathbf{M}_{k}^{H}\mathbf{V})}^{-1}},  &  \label{LDLinvFreeNov17b}
\end{subnumcases}
where  the diagonal $\mathbf{G}$,  the  upper-triangular $\mathbf{F}$ and ${{\mathbf{E}}}$ are the inverse
$\mathbf{LD}\mathbf{M}^H$  factors of $\mathbf{T}-{{\mathbf{Y}}^{H}}\mathbf{L}_{k}^{{}}\mathbf{D}_{k}^{{}}\mathbf{M}_{k}^{H}\mathbf{V}$ in (\ref{LDLinvFreeNov17b}).
The derivation of
(\ref{ABFGEcomputeE84r9834}) is as follows.

From (\ref{LDMdefineAll313943}) we can deduce
$\mathbf{L}_{k+i}^{-1}\text{=}\left[ \begin{matrix}
   \mathbf{L}_{k}^{-1} & \text{-}\mathbf{L}_{k}^{-1}\mathbf{A}{{\mathbf{F}}^{-1}}  \\
   \mathbf{0} & {{\mathbf{F}}^{-1}}  \\
\end{matrix} \right]$,
$\mathbf{D}_{k+i}^{-1}=\left[ \begin{matrix}
   \mathbf{D}_{k}^{-1} & \mathbf{0}  \\
   \mathbf{0} & {{\mathbf{G}}^{-1}}  \\
\end{matrix} \right]$
and
$\mathbf{M}_{k+i}^{-1}\text{=}\left[ \begin{matrix}
   \mathbf{M}_{k}^{-1} & \text{-}\mathbf{M}_{k}^{-1}\mathbf{B}{{\mathbf{E}}^{-1}}  \\
   \mathbf{0} & {{\mathbf{E}}^{-1}}  \\
\end{matrix} \right]$,
which are substituted into (\ref{MinvDL2Re4332}) to obtain
\begin{multline}\label{LDLinvFreeNov17a}
\left[ \begin{matrix}
   \mathbf{M}_{k}^{-H} & \mathbf{0}  \\
   -{{\mathbf{E}}^{-H}}{{\mathbf{B}}^{H}}\mathbf{M}_{k}^{-H} & {{\mathbf{E}}^{-H}}  \\
\end{matrix} \right]\left[ \begin{matrix}
   \mathbf{D}_{k}^{-1} & \mathbf{0}  \\
   \mathbf{0} & {{\mathbf{G}}^{-1}}  \\
\end{matrix} \right]  \\
\times
\left[ \begin{matrix}
   \mathbf{L}_{k}^{-1} & -\mathbf{L}_{k}^{-1}\mathbf{A}{{\mathbf{F}}^{-1}}  \\
   \mathbf{0} & {{\mathbf{F}}^{-1}}  \\
\end{matrix} \right]=\mathbf{R}_{k+i}^{{}}.
 \end{multline}
 We can substitute (\ref{R2RVT324913}) into (\ref{LDLinvFreeNov17a}) to  deduce
\begin{multline}\label{}
\left[ {\begin{array}{*{20}{c}}
{{\bf{M}}_k^{ - H}{\bf{D}}_k^{ - 1}{\bf{L}}_k^{ - 1}}&{ - {\bf{M}}_k^{ - H}{\bf{D}}_k^{ - 1}{\bf{L}}_k^{ - 1}{\bf{A}}{{\bf{F}}^{ - 1}}}\\
{\left( \begin{array}{l}
{\rm{ - }}{{\bf{E}}^{ - H}} \times \\
{{\bf{B}}^H}{\bf{M}}_k^{ - H} \times \\
{\bf{D}}_k^{ - 1}{\bf{L}}_k^{ - 1}
\end{array} \right)}&{\left( \begin{array}{l}
{{\bf{E}}^{ - H}}{{\bf{B}}^H}{\bf{M}}_k^{ - H} \times \\
{\bf{D}}_k^{ - 1}{\bf{L}}_k^{ - 1}{\bf{A}}{{\bf{F}}^{ - 1}} + \\
{{\bf{E}}^{ - H}}{{\bf{G}}^{ - 1}}{{\bf{F}}^{ - 1}}
\end{array} \right)}
\end{array}} \right] \\
 = \left[ {\begin{array}{*{20}{c}}
{{{\bf{R}}_k}}&{\bf{V}}\\
{{{\bf{Y}}^H}}&{\bf{T}}
\end{array}} \right],
\end{multline}
from which we can obtain
\begin{subnumcases}{\label{MEYEBMtotal312405}}
-\mathbf{M}_{k}^{{\text{-}}H}\mathbf{D}_{k}^{{\text{-}}1}\mathbf{L}_{k}^{{\text{-}}1}\mathbf{A}{{\mathbf{F}}^{{\text{-}}1}}=\mathbf{V} &  \label{MDLAF2V32143}\\
-{{\mathbf{E}}^{{\text{-}}H}}{{\mathbf{B}}^{H}}\mathbf{M}_{k}^{{\text{-}}H}\mathbf{D}_{k}^{{\text{-}}1}\mathbf{L}_{k}^{{\text{-}}1}={{\mathbf{Y}}^{H}}  &  \label{EBMDL2Y54989}  \\
{{\mathbf{E}}^{{\text{-}}H}}{{\mathbf{B}}^{H}}\mathbf{M}_{k}^{{\text{-}}H}\mathbf{D}_{k}^{{\text{-}}1}\mathbf{L}_{k}^{{\text{-}}1}\mathbf{A}
{{\mathbf{F}}^{{\text{-}}1}}+{{\mathbf{E}}^{{\text{-}}H}}{{\mathbf{G}}^{{\text{-}}1}}{{\mathbf{F}}^{{\text{-}}1}}=\mathbf{T}.    &  \label{EBMDLAFEGF2T438}
\end{subnumcases}

From  (\ref{MDLAF2V32143})  and (\ref{EBMDL2Y54989}) ,  we can deduce (\ref{A2LFMVF8430})   and (\ref{B2EYLDM3243})  ,  respectively. Then we substitute (\ref{A2LFMVF8430})  and (\ref{B2EYLDM3243})  into  (\ref{EBMDLAFEGF2T438})
 to obtain
${{\mathbf{Y}}^{H}}\mathbf{L}_{k}^{{}}\mathbf{D}_{k}^{{}}\mathbf{M}_{k}^{H}\mathbf{V}+{{\mathbf{E}}^{-H}}{{\mathbf{G}}^{-1}}{{\mathbf{F}}^{-1}}=\mathbf{T}$, from which we can deduce (\ref{LDLinvFreeNov17b}) .

\subsection{Inverse LU Factorization of a Partitioned Matrix}

Substitute (\ref{ABFGEcomputeE84r9834})
into
(\ref{LDMdefineAll313943}),
and substitute  (\ref{LDMdefineAll313943})
into (\ref{LDMtFactorsDefine398da32})  to obtain
\begin{multline}\label{LDLinvFreeNov17gOriginal}
\left[ \begin{matrix}
   {{\mathbf{L}}_{k}} & -\mathbf{L}_{k}^{{}}\mathbf{D}_{k}^{{}}\mathbf{M}_{k}^{H}\mathbf{VF}  \\
   \mathbf{0} & \mathbf{F}  \\
\end{matrix} \right]  \left[ \begin{matrix}
   {{\mathbf{D}}_{k}} & \mathbf{0}  \\
   \mathbf{0} & \mathbf{G}  \\
\end{matrix} \right] \times  \\
\left[ \begin{matrix}
   {{\mathbf{M}}_{k}^H} &  \mathbf{0}  \\
 -{{\mathbf{E}}^{H}}{{\mathbf{Y}}^{H}}\mathbf{L}_{k}^{{}}\mathbf{D}_{k}^{{}}\mathbf{M}_{k}^{H}   & \mathbf{E}^H  \\
\end{matrix} \right]
=\mathbf{R}_{k+i}^{-1}.
 \end{multline}
 From (\ref{LDLinvFreeNov17gOriginal})  we can deduce
\begin{multline}\label{LUinvFreeNov17a329}
\left[ \begin{matrix}
   {{\mathbf{L}}_{k}} & -{{\mathbf{L}}_{k}}{{\mathbf{D}}_{k}}{{\mathbf{M}}_{k}^{H}} \mathbf{VF}  \\
   \mathbf{0} & \mathbf{F}  \\
\end{matrix} \right]\left[ \begin{matrix}
   {{\mathbf{D}}_{k}} {{\mathbf{M}}_{k}^{H}}  & \mathbf{0}  \\
   -\mathbf{G}{{\mathbf{E}}^{H}}{{\mathbf{Y}}^{H}}{{\mathbf{L}}_{k}}{{\mathbf{D}}_{k}}{{\mathbf{M}}_{k}^{H}}  & \mathbf{G}{{\mathbf{E}}^{H}}  \\
\end{matrix} \right]  \\
=\mathbf{R}_{k+i}^{-1}.
 \end{multline}
Let ${{\mathbf{U}}_{k}}={{\mathbf{D}}_{k}}\mathbf{M}_{k}^{H}$,  $\mathbf{P}=\mathbf{G}{{\mathbf{E}}^{H}}$, which can be substituted into
 (\ref{LUinvFreeNov17a329})  and  (\ref{LDLinvFreeNov17b})  to obtain
   \begin{equation}\label{LUgeneral948422}
\left[ \begin{matrix}
   {{\mathbf{L}}_{k}} & -{{\mathbf{L}}_{k}}{{\mathbf{U}}_{k}}\mathbf{VF}  \\
   \mathbf{0} & \mathbf{F}  \\
\end{matrix} \right]\left[ \begin{matrix}
   {{\mathbf{U}}_{k}} & \mathbf{0}  \\
   -\mathbf{P}{{\mathbf{Y}}^{H}}{{\mathbf{L}}_{k}}{{\mathbf{U}}_{k}} & \mathbf{P}  \\
\end{matrix} \right]=\mathbf{R}_{k+i}^{-1}
 \end{equation}
and
  \begin{equation}\label{FPdefine3981083}
\mathbf{FP}={{(\mathbf{T}-{{\mathbf{Y}}^{H}}{{\mathbf{L}}_{k}}{{\mathbf{U}}_{k}}\mathbf{V})}^{-1}},
 \end{equation}
respectively. From (\ref{LUgeneral948422}) we can deduce
\begin{subnumcases}{\label{}}
{{{{\mathbf{L}}_{k+i}}}}=\left[ \begin{matrix}
   {{\mathbf{L}}_{k}} & -{{\mathbf{L}}_{k}}{{\mathbf{U}}_{k}}\mathbf{VF}  \\
   \mathbf{0} & \mathbf{F}  \\
\end{matrix} \right] &  \label{}\\
 {{\mathbf{U}}_{k+i}}=\left[ \begin{matrix}
   {{\mathbf{U}}_{k}} & \mathbf{0}  \\
   -\mathbf{P}{{\mathbf{Y}}^{H}}{{\mathbf{L}}_{k}}{{\mathbf{U}}_{k}} & \mathbf{P}  \\
\end{matrix} \right], &  \label{}
\end{subnumcases}
where $\mathbf{F}$ and $\mathbf{P}$ are the inverse LU factors satisfying
(\ref{FPdefine3981083}).

\subsection{Division Free Inverse LDM$^T$ Factorization}
Let us try to
use the  alternative
 LDM$^T$ factors of ${{\mathbf{R}}_{k}^{-1}}={\mathbf{Q}}_{k}$, which are assumed to be
    \begin{equation}\label{LDLinvFreeNov17c}
{\mathbf{\tilde{L}}}_{k}   ({\mathbf{\tilde{D}}}_{k}   /{\delta}_{k}    ){{\mathbf{\tilde{M}}}_{k}^{H}}   ={\mathbf{L}}_k {\mathbf{D}}_k{{\mathbf{M}}_k^H} ={{\mathbf{R}}_k^{-1}}={{\mathbf{Q}}_k}.
 \end{equation}
      Substitute
   (\ref{LDLinvFreeNov17c})   into (\ref{LDLinvFreeNov17b})   to obtain
$\mathbf{FG}{{\mathbf{E}}^{H}}={{\left( \mathbf{T}-{{\mathbf{Y}}^{H}}{\mathbf{\tilde{L}}}_{k}   ({\mathbf{\tilde{D}}}_{k}   /{\delta}_{k}    ){{\mathbf{\tilde{M}}}_{k}^{H}}   \mathbf{V} \right)}^{-1}}$, i.e.,
    \begin{equation}\label{LDLinvFreeNov17d}
\mathbf{F}(\mathbf{G}/{\delta}_{k}    ){{\mathbf{E}}^{H}}={{\left( {\delta}_{k}    \mathbf{T}-{{\mathbf{Y}}^{H}}{\mathbf{\tilde{L}}_k}{\mathbf{\tilde{D}}_k}{{\mathbf{\tilde{M}}}_{k}^{H}}   \mathbf{V} \right)}^{-1}}.
 \end{equation}
Assume that the division free LDM$^T$ factorization (\ref{LDLinvFreeNov17c}) is also utilized to obtain
    \begin{equation}\label{LDLinvFreeNov17e}
\mathbf{\tilde{F}}(\mathbf{\tilde{G}}/\eta ){{\mathbf{\tilde{E}}}^{H}}={{\left( {\delta}_{k}    \mathbf{T}-{{\mathbf{Y}}^{H}}{\mathbf{\tilde{L}}_k}{\mathbf{\tilde{D}}_k}{{\mathbf{\tilde{M}}}_{k}^{H}}   \mathbf{V} \right)}^{-1}}.
 \end{equation}
By comparing (\ref{LDLinvFreeNov17e})  and (\ref{LDLinvFreeNov17d}) , we can deduce
    \begin{equation}\label{LDLinvFreeNov17f}
\mathbf{\tilde{F}}({\delta}_{k}    \mathbf{\tilde{G}}/\eta ){{\mathbf{\tilde{E}}}^{H}}=\mathbf{FG}{{\mathbf{E}}^{H}}.
 \end{equation}

From (\ref{LDLinvFreeNov17c})  and (\ref{LDLinvFreeNov17f}) , we can write $\frac{{\mathbf{\tilde{D}}}_{k}   }{{\delta}_{k}   } ={\mathbf{D}}_k$, ${\mathbf{\tilde{L}}}_{k}   ={\mathbf{L}}_k$,
 ${{\mathbf{\tilde{M}}}_k}={{{\mathbf{M}}_k}}$,  $\frac{{\delta}_{k}   }{\eta} \mathbf{\tilde{G}} =\mathbf{G}$,  $\mathbf{\tilde{F}}=\mathbf{F}$ and  ${{\mathbf{\tilde{E}}}^{H}}={{\mathbf{E}}^{H}}$,
which are substituted into
(\ref{LDLinvFreeNov17gOriginal})
  to obtain
\begin{multline}\label{LDLinvFreeNov17g}
\left[ \begin{matrix}
   {{\mathbf{\tilde{L}}}_{k}   } & -{\mathbf{\tilde{L}}}_{k}   \frac{{\mathbf{\tilde{D}}}_{k}   }{{\delta}_{k}   }{{\mathbf{\tilde{M}}}_{k}^{H}}   \mathbf{V\tilde{F}}  \\
   \mathbf{0} & {\mathbf{\tilde{F}}}  \\
\end{matrix} \right]\left[ \begin{matrix}
  \frac{{\mathbf{\tilde{D}}}_{k}   }{{\delta}_{k}   }   & \mathbf{0}  \\
   \mathbf{0} & \frac{{\delta}_{k}   }{\eta} \mathbf{\tilde{G}}   \\
\end{matrix} \right] \times  \\
 \left[ \begin{matrix}
   {{\mathbf{\tilde{M}}}_{k}^{H}}    & \mathbf{0}  \\
   -{{{\mathbf{\tilde{E}}}}^{H}} {{\mathbf{Y}}^{H}}{\mathbf{\tilde{L}}}_{k}   \frac{{\mathbf{\tilde{D}}}_{k}   }{{\delta}_{k}   }{{\mathbf{\tilde{M}}}_{k}^{H}}    & {{{\mathbf{\tilde{E}}}}^{H}}  \\
\end{matrix} \right]
=\mathbf{R}_{k+i}^{-1}.
 \end{multline}
 To verify (\ref{LDLinvFreeNov17g}),
 we only need to
  substitute (\ref{LDLinvFreeNov17c})  and (\ref{LDLinvFreeNov17f})  into the left side of (\ref{LDLinvFreeNov17g}) (with the matrix multiplications finished), to verify that it is equal to the
 left side of (\ref{LDLinvFreeNov17gOriginal}) (with the matrix multiplications finished).

 From (\ref{LDLinvFreeNov17g})  we can deduce
\begin{multline}\label{LDLinvFreeNov17h}
\left[ \begin{matrix}
   {{\mathbf{\tilde{L}}}_{k}   } & -{\mathbf{\tilde{L}}_k}{\mathbf{\tilde{D}}_k}{{\mathbf{\tilde{M}}}_{k}^{H}}   \mathbf{V\tilde{F}}  \\
   \mathbf{0} & {\delta}_{k}    \mathbf{\tilde{F}}  \\
\end{matrix} \right]\left[ \begin{matrix}
 \frac{{\mathbf{\tilde{D}}}_{k}}{{\delta}_{k}   }    & \mathbf{0}  \\
   \mathbf{0} & \frac{\mathbf{\tilde{G}}}{{\delta}_{k}    \eta}  \\
\end{matrix} \right] \times  \\
\left[ \begin{matrix}
   {{\mathbf{\tilde{M}}}_{k}^{H}}    & \mathbf{0}  \\
   -{{\mathbf{Y}}^{H}}{\mathbf{\tilde{L}}_k}{\mathbf{\tilde{D}}_k}{{\mathbf{\tilde{M}}}_{k}^{H}}    & {\delta}_{k}    {{{\mathbf{\tilde{E}}}}^{H}}  \\
\end{matrix} \right]
=\mathbf{R}_{k+i}^{-1}.
 \end{multline}
Finally from (\ref{LDLinvFreeNov17h}),  we can obtain
\begin{subnumcases}{\label{GeneralEquGroup131}}
{{\mathbf{\tilde{L}}}_{k+i}}=\left[ \begin{matrix}
   {{{\mathbf{\tilde{L}}}}_{k}} & -{{{\mathbf{\tilde{L}}}}_{k}}{{{\mathbf{\tilde{D}}}}_{k}}\mathbf{\tilde{M}}_{k}^{H}\mathbf{V\tilde{F}}  \\
   \mathbf{0} & {\delta}_k \mathbf{\tilde{F}}  \\
\end{matrix} \right] &  \label{}\\
\mathbf{\tilde{M}}_{k+i}^{H}=\left[ \begin{matrix}
   \mathbf{\tilde{M}}_{k}^{H} & \mathbf{0}  \\
   -{{\mathbf{Y}}^{H}}{{{\mathbf{\tilde{L}}}}_{k}}{{{\mathbf{\tilde{D}}}}_{k}}\mathbf{\tilde{M}}_{k}^{H} & {\delta}_k {{{\mathbf{\tilde{E}}}}^{H}}  \\
\end{matrix} \right]  &  \label{}\\
{{\mathbf{\tilde{D}}}_{k+i}}=\left[ \begin{matrix}
   \eta {{{\mathbf{\tilde{D}}}}_{k}} & \mathbf{0}  \\
   \mathbf{0} & {\mathbf{\tilde{G}}}  \\
\end{matrix} \right]   &  \label{etaDcompute39212}\\
 {{\delta }_{k+i}}={{\delta }_{k}}\eta, &  \label{}
\end{subnumcases}
where $\mathbf{\tilde{F}}$,  $\mathbf{\tilde{G}}$,  ${{\mathbf{\tilde{E}}}}$ and $\eta$  are the division free LDM$^T$ factors computed by (\ref{LDLinvFreeNov17e}).

\section{Square-root and Division Free Inverse LDL$^T$ Factorization for V-BLAST}

When $i=1$, $\mathbf{V}$, ${{\mathbf{Y}}}$ and ${{\mathbf{T}}}$ in (\ref{R2RVT324913}) and (\ref{GeneralEquGroup131})
can be written as ${{\mathbf{v}}}$, $\mathbf{y}$ and $t$, respectively.
We can use (\ref{LDLinvFreeNov17e})  to obtain
\begin{subnumcases}{\label{LDLinvFreeNov17iAddVecGroup}}
\mathbf{\tilde{F}}=\mathbf{\tilde{G}}={{\mathbf{\tilde{E}}}^{H}}=1 &  \label{}\\
\eta =1/({{\delta }_{k}}t-\mathbf{y}^{H}{\mathbf{\tilde{L}}_k}{\mathbf{\tilde{D}}_k}{{\mathbf{\tilde{M}}}_k^{H}}{{\mathbf{v}}}). &  \label{}
\end{subnumcases}
Then we can use (\ref{LDLinvFreeNov17iAddVecGroup}) and (\ref{GeneralEquGroup131}) to compute
${{\mathbf{\tilde{L}}}_{k+1}}$,   ${{\mathbf{\tilde{D}}}_{k+1}}$, ${{\mathbf{\tilde{M}}}_{k+1}}$
and ${{\delta }_{k+1}}$ from ${{\mathbf{\tilde{L}}}_{k}}$,   ${{\mathbf{\tilde{D}}}_{k}}$, ${{\mathbf{\tilde{M}}}_{k}}$
and ${{\delta }_{k}}$ iteratively, and the iterations can start from the initial
\begin{subnumcases}{\label{S16NoDIVkOne}}
{\bf{{\tilde{L}}}}_{1}={\bf{{\tilde{M}}}}_{1}=1, & \label{S16NoDIVkOne1}\\
{\bf{{\tilde{D}}}}_{1}=1, & \label{S16NoDIVkOneD}\\
{\delta _{1}}={\bf{R}}_{1}=r_{1,1}, & \label{S16NoDIVkOne2}
\end{subnumcases}
where $r_{1,1}$  is the entry in the $1^{st}$ row and column of ${\bf{R}}$.

The iterations in (\ref{GeneralEquGroup131}) will lead to numerically unlimited
results, which may cause a problem in fixed-point implementations
\cite{WCNCcholesky}. We can alleviate this problem by scaling, as in
\cite{WCNCcholesky}. Scaling is achieved by dividing (or
multiplying) only by powers of $2$ \cite{WCNCcholesky}, which is a
shift operation in binary fixed-point implementation. Since ${\delta
_{k}}$ is complex, we can keep $\left| {\delta _{k}} \right|^2$
between $0.25$ and $4$, and scale the diagonal entries in
${\bf{\tilde D}}_{k}$ accordingly. Thus ${\delta _{k}}$ and ${\bf{\tilde D}}_{k}$
in (\ref{S16NoDIVkOne}) or (\ref{GeneralEquGroup131})
 are multiplied by $c_k$, which is a power
of $2$. Correspondingly in each iteration we end up with ${\delta
_{k}}c_k$ and ${\bf{\tilde D}}_{k}c_k$, while $\left|{\delta
_{k}}c_k\right|^2$ is always between $0.25$ and $4$.

Now we can apply
(\ref{S16NoDIVkOne}), (\ref{LDLinvFreeNov17iAddVecGroup}) and (\ref{GeneralEquGroup131})
to compute
 the alternative
LDL$^T$ factors
of
the Hermitian
${\bf{{{Q}}}}$ in (\ref{equ:4}), i.e., ${\bf{L}}_{K}$,  ${\bf{D}}_{K}$ and ${\delta
_{K}}$ satisfying
\begin{equation}\label{S28:DefLDLdetaofQ}
{\bf{L}}_{K}({\bf{D}}_{K}/{\delta
_{K}}){\bf{L}}_{K}^H={\bf{Q}}_{K}={\bf{R}}_{K}^{ - 1}.
\end{equation}
The initial ${\bf{L}}_{|K}={\bf{L}}_{K}$,
${\bf{D}}_{|K}={\bf{D}}_{K}$ and $\delta=\delta_{K}$ are obtained
 after $K-1$ iterations, which start from ${\bf{L}}_{1}$, ${\bf{D}}_{1}$ and $\delta _{1}$
 in (\ref{S16NoDIVkOne}). 

 The proposed recursive V-BLAST algorithm
 computes the initial ${\bf{Q}}_{|K}={\bf{Q}}_{K}$ by (\ref{S28:DefLDLdetaofQ}),
 and then computes ${\bf{Q}}_{|(k-1)}$ from ${\bf{Q}}_{|k}$ ($k=K,K-1,\cdots,2$) in the OSIC detection
phase by the
 recursive algorithm proposed in \cite{TransWC2009ReCursiveBlast}.

With the initial ${\bf{L}}_{|K}$,
${\bf{D}}_{|K}$ and $\delta$, we can also
 propose a square-root and division free implementation of the square-root V-BLAST algorithm
 in \cite{zhfvtc2008}.  In the $k^{th}$ ($k=K,K-1,\cdots,2$) iteration of the OSIC detection,
 we find  a non-unitary transformation ${\bf{\Theta  }}$ and the corresponding
diagonal ${\bf{{D'} }}_{|k}$ that satisfy
 \begin{equation}\label{S18:AftNOFinalUpdatLD}
({\bf{L}}_{|k} {\bf{\Theta  }}){\bf{{D'} }}_{|k}
\left({\bf{L}}_{|k} {\bf{\Theta  }}\right)^H ={\bf{L}}_{|k}{\bf{D}}_{|k}
{\bf{L}}_{|k}^H,
\end{equation}
 where ${\bf{\Theta  }}$ block upper-triangularizes ${\bf{L}}_{|k}$, i.e.,
 \begin{equation}\label{S27TriangleL}
{\bf{L}}_{|k} {\bf{\Theta  }} = \left[ {\begin{array}{*{20}c}
   {{\bf{L}}_{|k-1} } & {{\bf{\mu }}_{k-1}}  \\
   {{\bf{0}}_{k-1}^T } & {\lambda _k}  \\
\end{array}} \right].
\end{equation}
In (\ref{S27TriangleL}),  ${\bf{\mu }}_{k - 1}$ and $\lambda_k$ denote a column vector and a
scalar, respectively, and the transformation ${\bf{\Theta  }}$ can be performed
by a series of wide-sense Givens rotations proposed in \cite{zhfVTC2010DivFree}, which are square-root and division free.
${\bf{L}}_{|k-1}$ for the next iteration is the sub-matrix in (\ref{S27TriangleL}), and
${\bf{D}}_{|k-1}$ for the next iteration is obtained by removing
the last row and column in ${\bf{{D'} }}_{|k}$.

%
%
%
%
%
%

\section{Complexity Evaluation}

As in \cite{zhfvtc2008}, let
 ($j$, $k$) denote the complexity of $j$ complex multiplications
and $k$ complex additions, and simplify ($j$, $k$) to ($j$) if
$j=k$.   The complexity to compute the initial ${\bf{L}}_{|K}$,
${\bf{D}}_{|K}$ and $\delta$
by (\ref{S16NoDIVkOne}), (\ref{LDLinvFreeNov17iAddVecGroup}) and (\ref{GeneralEquGroup131})
is $(\frac{1}{3} K^3)$,
which is about the same as the complexity of
the Cholesky factorization with the
back substitution~\cite{Matrix_Computations_book,FixedPointChskInv}.
Moreover, the complexity to
compute the initial ${\bf{Q}}_{|K}$ by (\ref{S28:DefLDLdetaofQ}) is $(\frac{1}{6} K^3)$.
Then the total complexity to compute the initial ${\bf{Q}}_{|K}$ is $(\frac{1}{2} K^3)$,
which is equal to the complexity of the recursive algorithm~\cite{TransWC2009ReCursiveBlast,zhfICC2009} to compute ${\bf{Q}}_{|K}$.
Accordingly with respect to the existing recursive V-BLAST algorithm~\cite{TransWC2009ReCursiveBlast,zhfICC2009},
the proposed recursive V-BLAST algorithm requires the same complexity, and saves $K-1$ divisions in the initial step
to compute ${\bf{Q}}_{|K}$.  Moreover, after a very large number of iterations,
the recursive algorithm to compute ${\bf{Q}}_{|K}$ may introduce numerical instabilities~\cite{zhf3} in the processor units with
the finite precision, while usually the ${\bf{LD}}{{\bf{L}}^T}$
   factorization is numerically stable~\cite{Matrix_Computations_book}.

 With respect to
the square-root V-BLAST algorithm in
\cite{zhfvtc2008},
the proposed
square-root and division free
 V-BLAST algorithm requires  about
 the same computational complexity,
 which  ranges from~\cite[Table \Rmnum{1}]{zhfvtc2008}
($\frac{2}{3}K^{3}+\frac{1}{2}K^{2}N,\frac{4}{9}K^{3}+\frac{1}{2}K^{2}N$)
to ($\frac{1}{3}K^{3}+\frac{1}{2}K^{2}N$).
To compute the initial square-root ${\bf{\Gamma }}$, we can also use the
alternative Cholesky factorization in \cite{WCNCcholesky} plus the
back substitution~\cite{Matrix_Computations_book,FixedPointChskInv}, which
 requires $\frac{1}{2}K^{3}+O(K^{2})$ more of real multiplications~\cite{WCNCcholesky}
 than the conventional Cholesky factorization with the
back substitution, and still requires $K$ divisions for the back substitution~\cite{Matrix_Computations_book,FixedPointChskInv}.
Moreover, the OSIC square-root V-BLAST algorithm
in \cite{BLASTtransSP2015} still utilizes the
conventional Cholesky factorization with the
back substitution~\cite{Matrix_Computations_book,FixedPointChskInv} to compute the initial square-root
${\bf{\Gamma }}$, and
 the improvement in \cite{BLASTtransSP2015} is that
 the back-substitution
reuses the results of the divisions in the Cholesky factorization, to reduce half divisions and compute
the initial square-root ${\bf{\Gamma }}$ by only $K$ divisions.
It can easily be seen that the OSIC V-BLAST algorithm in \cite{BLASTtransSP2015} requires the same complexity
as
the V-BLAST algorithm
in  \cite{zhfvtc2008}, and they both spend only $K$ divisions to compute the initial square-root ${\bf{\Gamma }}$.
Thus in the following Fig. 1, the V-BLAST algorithm in \cite{BLASTtransSP2015} is not  simulated.

%
%
%
%
%
%
%
%
%
%
%
%

For different number of transmit/receive antennas,
 some numerical experiments were carried out to count the average flops of
the presented algorithms. The results are shown in Fig. 1. It can be
seen that they are consistent with the theoretical flops
calculation.



\begin{figure}[htbp]
\centering
\includegraphics[height=4.5cm, width=7.5cm]{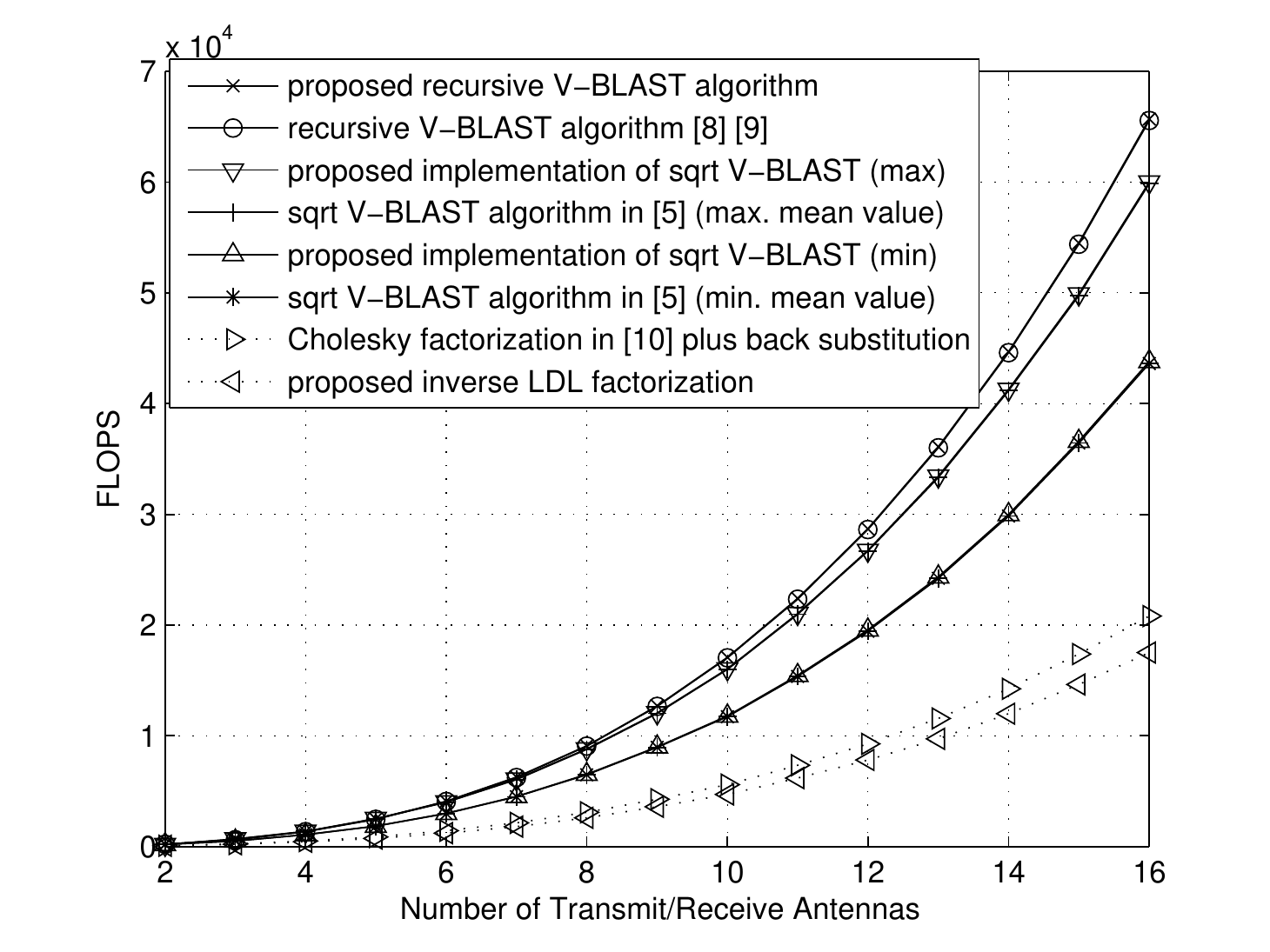}
\caption{Complexity Comparison among the Presented Algorithms.}
\end{figure}

\section{Conclusion}

In this letter,  the inverse LDM$^T$ and LU factorizations are proposed
for a matrix partitioned into $2 \times 2$ blocks, which include the square-root and division free version.
The proposed square-root and division free inverse LDM$^T$ factorization
is  applied to compute the initial
   estimation error covariance matrix
    $\mathbf{Q}$ for the recursive  V-BLAST algorithm,
    and is also applied to propose the square-root and division free implementation for the square-root V-BLAST algorithm in \cite{zhfvtc2008}, where the
 wide-sense Givens rotation in \cite{zhfVTC2010DivFree} is utilized.
 With respect to the existing recursive V-BLAST algorithm~\cite{TransWC2009ReCursiveBlast,zhfICC2009},
 the recursive  V-BLAST algorithm with the initial $\mathbf{Q}$ computed by the proposed square-root and division free
  inverse LDM$^T$ factorization requires about the same computational complexity, and can save $K-1$ divisions.
  With respect to the  existing square-root V-BLAST algorithms~\cite{zhfvtc2008,BLASTtransSP2015}, the proposed
  square-root and division free V-BLAST algorithm requires about the same computational complexity, and can avoid the
  square-root and division operations.


\begin{thebibliography}{34}

\bibitem{MIMOintroduce} G. J. Foschini and M. J. Gans,
``On limits of wireless communications in a fading environment
when using multiple antennas", \emph{Wireless Personal
Commun.}, pp.
311-335, Mar. 1998.

\bibitem{zhf1} P. W. Wolniansky, G. J. Foschini, G. D. Golden and R. A. Valenzuela,
``V-BLAST: an architecture for realizing very high data rates over
the rich-scattering wireless channel", \emph{Proc. ISSSE 98}, pp.
295-300, 1998.




\bibitem{zhf2} B. Hassibi, ``An efficient square-root algorithm for BLAST",
\emph{IEEE ICASSP '00}, pp. 737-740, June 2000.

\bibitem{zhfsqrtSPletters} H. Zhu, Z. Lei, and F. P. S. Chin, ``An improved square-root algorithm
for BLAST", \emph{IEEE Signal Process. Lett.}, vol. 11, no. 9, pp. 772-775, Sep.
2004.

\bibitem{zhfvtc2008} H. Zhu, W. Chen, B. Li, and F. Gao, ``An Improved Square-Root Algorithm for V-BLAST Based on
Efficient Inverse Cholesky Factorization", \emph{IEEE Trans. Wireless Commun.}, vol. 10,
no. 1, Jan. 2011.

\bibitem{BLASTtransSP2015} K. Pham and K. Lee,  ``Low-Complexity SIC Detection Algorithms for
 Multiple-Input Multiple-Output Systems", \emph{IEEE
Trans. on Signal Processing}, pp. 4625-4633, vol. 63, no. 17, Sept. 2015.


\bibitem{zhf3} J. Benesty, Y. Huang and J. Chen, ``A fast recursive algorithm for
optimum sequential signal detection in a BLAST system", \emph{IEEE
Trans. on Signal Processing}, pp. 1722-1730, July 2003.



 \bibitem{TransWC2009ReCursiveBlast}  Y. Shang and X. G. Xia, ``On fast recursive algorithms for V-BLAST
with optimal ordered SIC detection", \emph{IEEE Trans. Wireless Commun.}, vol. 8, pp. 2860-2865, June 2009.

\bibitem{zhfICC2009} H. Zhu, W. Chen and F. She, ``Improved Fast Recursive Algorithms for
V-BLAST and G-STBC with Novel Efficient Matrix Inversion,"
\emph{IEEE ICC 2009}, Dresden, Germany, June 2009.

\bibitem{WCNCcholesky}
L. M. Davis, ``Scaled and decoupled Cholesky and QR decompositions
with application to spherical MIMO detection", \emph{IEEE WCNC,
2003}.  

\bibitem{kjLiuGIVENS}
E. N. Frantzeskakis and K. J. R. Liu, ``A class of square root and
division free algorithms and architectures for QRD-based adaptive
signal processing", \emph{IEEE Trans. on Signal Processing}, Sep
1994.

\bibitem{Matrix_Computations_book}
 G. H. Golub and C. F. Van Loan, \emph{Matrix Computations}, Johns Hopkins University Press,
 Baltimore, MD, 3rd edition, 1996.

 \bibitem{Matlab_inv_function_introduce} https://ww2.mathworks.cn/help/matlab/ref/inv.html?lang=en.


\bibitem{zhfVTC2010DivFree} H. Zhu, W. Chen, and B. Li, ``Efficient Square-Root and Division Free Algorithms for Inverse $LDL^T$ Factorization and the Wide-Sense Givens Rotation with
Application to V-BLAST", \emph{IEEE Vehicular Technology Conference (VTC)}, 2010
Fall, 6-9 Sept., 2010.






 \bibitem{FixedPointChskInv}
A. Burian, J. Takala, M. Ylinen, ``A fixed-point implementation of
matrix inversion using Cholesky decomposition",
\emph{IEEE International Symposium on MHS}, Dec. 2003, Vol. 3, pp. 1431-1434.


%
%




\end{thebibliography}
\end{document}